# The "Signalgate" Case is Waving a Red Flag to All Organizational and Behavioral Cybersecurity Leaders, Practitioners, and Researchers: Are We Receiving the "Signal" Amidst the Noise?

**Copyright © 2025 Paul Benjamin Lowry** (Eminent Scholar and Suzanne Parker Thornhill Chair Professor, Pamplin College Business, Virginia Tech)**; Gregory D. Moody** (Lee Professor of Information Systems at the College of Business at UNLV); **Robert Willison** (Professor, Department of Intelligent Operations and Marketing, Suzhou Xi'an Jiaotong Liverpool University), and **Clay Posey** (Professor of Information Systems, Marriott School of Business, Brigham Young University).

## Abstract

The Signalgate incident of March 2025, wherein senior US national security officials inadvertently disclosed sensitive military operational details via the encrypted messaging platform Signal, highlights critical vulnerabilities in organizational security arising from human error, governance gaps, and the misuse of technology. Although smaller in scale when compared to historical breaches involving billions of records, Signalgate illustrates critical systemic issues often overshadowed by a focus on external cyber threats. Employing a case-study approach and systematic review grounded in the NIST Cybersecurity Framework (CSF), we analyze the incident to identify patterns of human-centric vulnerabilities and governance challenges common to organizational security failures. Findings emphasize three critical points: (1) organizational security depends heavily on human behavior, with internal actors often serve as the weakest link despite advanced technical defenses; (2) leadership's tone strongly influences organizational security culture and efficacy; and (3) widespread reliance on technical solutions without sufficient investments in human and organizational factors leads to ineffective practices and wasted resources. From these observations, we propose actionable recommendations for enhancing organizational and national security, including strong leadership engagement, comprehensive adoption of zero-trust architectures, clearer accountability structures, incentivized security behaviors, and rigorous oversight. Particularly during periods of organizational transition—such as mergers or large-scale personnel changes—additional measures become particularly important. In conclusion, Signalgate underscores the need for leaders and policymakers to reorient cybersecurity strategies toward addressing governance, cultural, and behavioral risks. Failure to do so perpetuates costly security failures and intensifies global vulnerabilities in a fragile security environment.

## 1. Introduction

In March 2025, the "Signalgate" incident erupted when it was revealed that top United States (US) national security officials had inadvertently shared sensitive military plans and actions for airstrikes in Yemen over a top-level group chat, including the Vice President of the US and other top leaders, on the encrypted messaging platform Signal. The group, intended for high-level coordination, mistakenly included a journalist, who later published the full details of the internal communications in The Atlantic. The leak disclosed precise operational timelines and targets, exposing critical gaps in communication security, compartmentalization, and situational awareness—even among senior officials in the US executive branch. Although the administration initially insisted no classified information was shared, the incident triggered bipartisan outrage, legal scrutiny, and a judicial preservation order, raising broader questions about the adequacy of existing security protocols and decision-making accountability at the highest levels of the US government.

Signalgate is among many significant security failures across public and private institutions worldwide. In fact, a litany of previous debacles eclipses it with respect to the number of records breached or underlying organizational damage, including but not limited to two breaches at Yahoo in 2013 (3 billion accounts) and 2014 (500 million accounts); in 2015 the US Office of Personnel Management was hacked, compromising highly private information of 22.1 million US Federal employees; in 2018, India's national ID database, Aadhaar was breached exposing 1.1 billion citizen accounts; in 2019 there was unauthorized access to 1.1 billion pieces of Alibaba user data; in 2020 the Cam4 adult streaming site accidentally exposed 10 billion private records due to database misconfiguration issues; in 2019, two datasets from Facebook were publicly exposed, comprising the privacy of 530 million users; in 2013, the Starwood reservation system acquired by Marriott was compromised and exposed private data of around 500 million guests; in 2013, Target's point-of-sales systems were hacked, exposing 40 million credit cards and 70 million customers' private data; in 2017 Equifax had a serious breach of around 145 million US citizen highly private data being compromised, including birth dates, social security numbers, addresses, and birthdates. These incidents resulted in multi-year investigations, negative political and social consequences, legal and government sanctions, costly civil lawsuits, and institutional reputational damage.

Considering these incidents, the Signalgate incident appears relatively benign; however, we assert it offers a rare, real-world lens into how organizational security can fail even at the highest organizational levels, not due to external cyberattacks but through human errors, flawed governance, and the misuse of technology assumed to be secure. We argue that the clear signals and patterns exemplified in Signalgate can be found in virtually all the other significant internal or external breaches. Signalgate exemplifies the convergence of behavioral risks, insider threat dynamics, and systemic lapses in organizational controls. It underscores the need to reexamine traditional security frameworks through the lens of human-centric vulnerabilities and governance culture. These issues make Signalgate an instructive case for scholars, practitioners, and policymakers seeking to understand and strengthen organizational security in an era where trust, technology, and transparency increasingly collide.

Signalgate is also critical because it is too easy to focus on the political scandal, interpret it through a partisan lens, and miss the more significant lesson that can be learned from the incident. From different political perspectives, some may emphasize the scandalous or embarrassing aspects, others may minimize or dispute its seriousness, and still others may tune it out entirely, given fatigue with the US political climate. However, to organizational leaders and practitioners, we have a different message: tune out the scandal itself, focus on the actual signal, and ignore the noise (e.g., politics, emotions, bias) to learn the essential lessons that can be gleaned from Signalgate.

As long-time organizational and behavioral security researchers who try to understand why people, organizations, and governments make poor security decisions, we see a consistent pattern in major breach cases. There are stark lessons to be learned if researchers and leaders only pay attention to the signal with humility, openness, and a willingness to learn, change, and lead by example. This pattern emerged yet again in the Signalgate event. Our goal, then, is to use this incident to underscore further how non-technical aspects can adversely affect institutional security even when the most stringent of policies are in place—something that continues to happen far too often, no matter the industry or region.

Because our aim is educational rather than political, we take this opportunity to emphasize several key points about our approach and findings. First, we have examined the incident using various sources to limit any potential biases in gathered information. Details of the incident used to arrive at our conclusions were open source and freely available as of **September 07, 2025**. Second, we publicized and made an initial version of this editorial, which was initially available to the general public on the arXiv repository before its submission for formal peer review, freely available to the general public on the arXiv repository. Comments from the general public to help clarify or correct our interpretation of the events related to the case have been incorporated, with formal acknowledgements provided at the end of the editorial. Third, the author team includes individuals from multiple political perspectives, living in and outside the US. Finally, this editorial should not be used to infer guilt for any criminal activity of any individuals named in the case or our analysis, nor should it imply any defect/inferiority of the Signal platform.

## 2. Summary of the Signalgate Case

The Signalgate incident emerged in March 2025. At its core, the event centers on a Signal group chat wherein information regarding military strikes on Houthi forces in Yemen was shared among the group members. U.S. National Security Advisor Mike Waltz created the group, which included US Vice President JD Vance, Secretary of Defense Pete Hegseth, Secretary of State Marco Rubio, CIA Director John Ratcliffe, Director of National Intelligence Tulsi Gabbard, and other cabinet-level national security officials. Due to Waltz's apparent mistake, the group included Jeffrey Goldberg, Editor-in-Chief of The Atlantic, who was not a government official and lacked security clearance. Over the course of five days, senior government officials exchanged what many intelligence professionals assessed to be highly sensitive, and in some instances top secret, operational military information—including the timing of airstrikes, weapons platforms, and target intelligence—in a format and setting that violated U.S. operational security standards and federal record-keeping laws. Again, based on public reporting, the event did not affect those military operations.

The case occurred within a broader context of significant personnel changes, which are typically expected and can be chaotic when a new administration takes the helm. However, upon returning to office in January 2025, President Trump's personnel changes were extensive and without precedent, including dismissing over a dozen Inspectors General, many of whom were responsible for ensuring accountability and oversight within security-sensitive agencies. Appointments such as Pete Hegseth to the position of Secretary of Defense drew scrutiny from some observers, who raised concerns about the vetting process and potential risks under federal personnel policies (e.g., Executive Order 10450). Though approved through formal and legal channels, key cabinet members, several of whom had little to no formal security training, were fast-tracked into roles with minimal formal security training, relying instead on statutory exceptions tied to their political appointments.

Using the encrypted Signal application contravened federal rules against employing ephemeral, unrecorded messaging platforms for official business involving sensitive or classified information. Subsequent reviews by military analysts and intelligence professionals revealed that the contents of the Signal chat met or exceeded the classification thresholds outlined in US federal statutes, intentional or otherwise. Some have argued that the event constituted a potential breach of the Espionage Act. Beyond the legal implications, reports indicate that the fallout extended to diplomatic consequences: it was later confirmed that Israeli-supplied human intelligence had been compromised in the chat, raising concerns among allies regarding the U.S. government's ability to safeguard shared intelligence in future operations.

Goldberg alerted the administration and the public on March 24, 2025, to the general information shared in the Signal chat. He published the full contents publicly on March 26, 2025, when administration officials wavered on the validity of his claims and the sensitivity of the details of the chat messages. Communication analysis revealed detailed pre-strike intelligence, operational sequencing, and internal coordination messages. At the behest of the CIA, Goldberg redacted the name of an undercover CIA officer, who also shared in the chat. As a result, public figures across the political spectrum, including Senate Armed Services Committee members, criticized the administration for undermining national security norms, and cybersecurity experts warned that the practices involved could have resulted in a significant loss of life had adversaries had access to the shared information.

In the following weeks, several institutional responses ensued: the Pentagon reissued guidance banning Signal for any operational or sensitive coordination, the NSA formally warned staff against third-party encrypted apps due to adversarial targeting, and select congressional committees opened preliminary inquiries. As of this publication, formal reprimands for the event have not been levied against specific individuals publicly, and the White House declared the matter "closed" on March 31. Interestingly, several individuals associated with Signalgate have been involved in employment changes. For example, General Timothy Haugh, the NSA Director, regarded by some as the most experienced and professionally qualified security official, was summarily fired in early April without explanation. Waltz was removed from his cabinet position and accepted a role at the UN. Whether these changes were in response to the event is unknown.

Signalgate illustrates what happens when organizational structures are in flux and institutional and security knowledge varies among members. Further, it serves as a case study in how technological tools, absent robust organizational safeguards, can completely undermine secure conduits for organizational communication, and thus is an archetypal demonstration that technological security methods can be limited and can always be thwarted by the humans involved. By bypassing controls through their actions, humans can circumvent technological controls; thus, security through technology alone will never really work. Moreover, Signalgate highlights critical intersections between institutional trust, the need to expeditiously deliver on an organization's operational and strategic goals, and the fragility of secure systems without robust controls or safeguards for the individuals using the technologies. For scholars and practitioners of organizational security, the event underscores the imperative to strengthen behavioral safeguards, reaffirm oversight institutions, and reimagine security policy as an ecosystem of practices based on Zero-trust architecture, where no one is "too senior" or "too powerful" for the same security governance and policies that apply to everyone else in the organization (e.g., Buck et al., 2021; Itodo & Ozer, 2024).

## 3. Analyzing Signalgate for the Missing "Signals" Using the NIST CSF Framework

The Signalgate incident is replete with missed signals and opportunities that could have prevented the incident in the first place. To determine the factors that likely led to the Signalgate incident, we will position individual components related to the complex case within the NIST cybersecurity framework (CSF).[1] NIST, a non-regulatory agency within the U.S. Department of Commerce, developed the CSF following Executive Order 13636 in 2013. It was designed with industry experts, government agencies, academia, and critical infrastructure stakeholders to ensure broad applicability and practicality. NIST CSF is widely adopted by organizations across industries and sectors globally, including:

- Private-sector businesses: to enhance cybersecurity resilience, manage risks, and align security practices with business objectives.
- Government entities: for compliance and standardized security practices in protecting critical infrastructure and sensitive data.
- Critical infrastructure sectors (e.g., energy, healthcare, financial services): to safeguard vital operations against disruption, enhance security oversight, and ensure regulatory compliance.
- International organizations: to standardize and align global cybersecurity efforts, thereby facilitating effective communication and coordination.

Thousands of firms worldwide use the NIST CSF framework to guide their cybersecurity efforts, including those that do not follow closely aligned frameworks. Thus, using NIST CSF provides an objective framework to analyze and discuss the Signalgate case and understand what can be learned in other organizational security settings. Global organizations extensively use NIST CSF for several reasons, because it:

- (1) Is adaptable to organizations of all sizes and industries;
- (2) Integrates well with other standards and compliance frameworks (e.g., ISO/IEC 27001, CMMC);
- (3) Provides a common nomenclature and systematic approach to addressing cybersecurity risks, facilitating communication between technical teams and executive leadership;
- (4) Enhances organizational security resilience by emphasizing continuous improvement and proactive risk management.

The NIST CSF 2.0 approaches organizational cybersecurity from six key perspectives: Identify, protect, detect, respond, recover, and govern, as depicted in Figure 1.

### 3.1. Identify

In the NIST CSF Framework 2.0, the **Identify** function aims to develop an organizational understanding of cybersecurity risks to systems, people, assets, data, and capabilities. This foundational function involves asset management, risk assessment, governance, and identifying critical infrastructure vulnerabilities. Key

[1] Like many analyses of major security cases in both public and private sectors, ours likely suffers from information asymmetry, where public knowledge is not a complete representation of the truth. Nonetheless, we believe that important lessons can be gleaned from what the public does know despite any potential inconsistencies or biases in available information.

**Figure 1**. The NIST Cybersecurity Framework 2.0 (CSF)

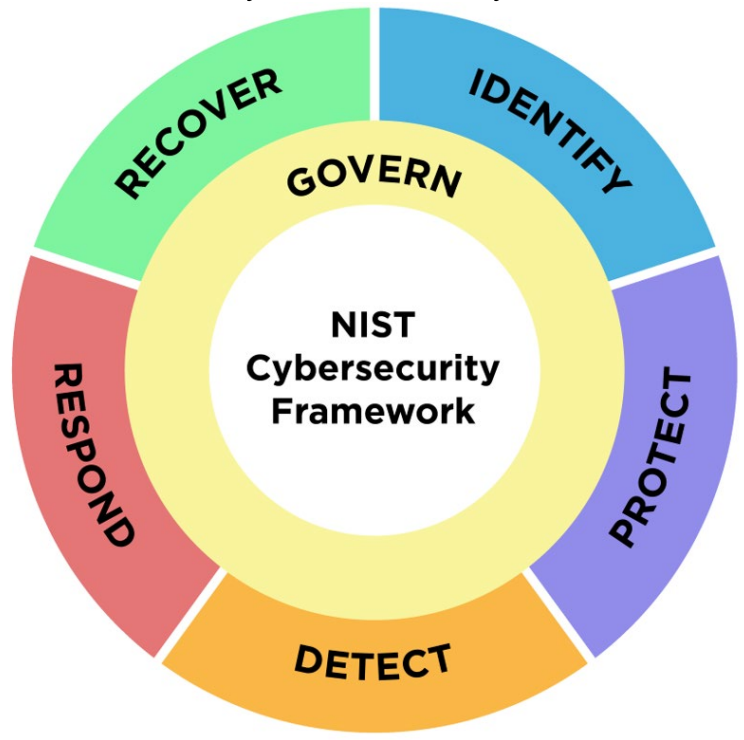

Note. Credit is to NIST and Natasha Hanacek

Signalgate issues regarding the Identify function include the following:

- Unclear roles and responsibilities for managing classified digital assets
- Failure to maintain consistent cybersecurity oversight during periods of significant organizational change or transformation
- Inadequate identification of critical communication assets, risks, and governance
- Poor inventory management of sensitive data assets requiring special handling
- Insufficient identification and classification of who should access sensitive communications
- Lack of understanding of sensitive, classified, or proprietary data
- Ignoring or not understanding compliance requirements related to sensitive information
- Inadequate cybersecurity governance and oversight

### 3.1.1. Unclear roles and responsibilities for managing classified digital assets

There was a breakdown in clearly defined roles and responsibilities for managing sensitive digital communications, including instances of unapproved platform use, such as Signal.[ID.GV-2] [1] This may have been exacerbated because many newly appointed individuals lacked documented responsibilities or cybersecurity training for classified asset management. Reports have suggested that numerous Signal chat groups were in use, often involving overlapping participants, making it challenging to maintain coordinated oversight or assign accountability within a structured governance framework.[2]

**Lessons:** Significant organizational change can substantially increase an organization's threat landscape (Herath & Rao, 2009; Kotulic & Clark, 2004). Data governance roles must be reassigned clearly and immediately during C-level executive transitions. Times of significant organizational change, such as mergers and acquisitions, corporate restructuring, and layoffs, should be considered times of internal and external security risk. In these times, roles and responsibilities are confused, internal struggles are more likely, employees are stressed, and are more prone to make bad decisions, especially when they are in new roles. Relevant to Signalgate, the DoD's *Behavioral Threat Analysis Center* (BTAC) issued a bulletin in 2022 warning of insider risks associated with major transitions, noting leadership changes, political environments, and related stresses as essential to monitor and mitigate.[3] Not surprisingly, multiple industry studies have pointed to a connection between major organizational changes, such as layoffs, restructuring, mergers, and acquisitions, and dramatically increasing internal and external cybersecurity risks.[4] Recent academic research has also found this tie with mergers and acquisitions (Liang et al., 2025).

### 3.1.2. Weak cybersecurity oversight during major organizational change or transformation

There was a failure to maintain consistent cybersecurity oversight of security governance during the transition to the new US administration.[ID.GV-2, ID.GV-3] These transitions are always challenging, but the new administration compounded continuity and oversight challenges by conducting a limited handoff process with the previous administration, which increased internal control of decisions but reduced external oversight. Sen. Elizabeth Warren warned in November 2024 that President-elect Trump's refusal to cooperate with standard transition protocols—such as signing Memoranda of Understanding—undermined the effectiveness and security of the handoff.[5] She warned that these deviations could **create national security vulnerabilities by disrupting established vetting and oversight processes.** Politico further reported in December 2024 that they were eschewing normal secure channels and were using email for much of the transition.[6]

They also fired approximately 17 US Inspector Generals (IGs), without the statutorily required 30-day notice to Congress. They act as independent watchdogs and are supposed to provide nonpartisan oversight of executive agencies and help ensure continuity and accountability across administrations.[7] These dismissals raised concerns about weakening institutional checks during increased scrutiny and transition. It has been argued that in several cases, experts were dismissed and replaced by individuals with limited prior expertise but with strong partisan alignment with the administration, and a couple with ties to the agencies they were to investigate, which further reduced professional oversight capacity and institutional knowledge.[8] If true, this action further hollowed out oversight because loyalty was placed at a premium over professional expertise and organizational learning.

**Lessons:** Abrupt dismissals or departures of key organizational members who are supposed to provide oversight, ethical leadership, and accountability increase the risk landscape for the organization (Harris et al., 2009).[9] Organizations must take their responsibilities regarding legally required oversight, independence, and reporting seriously and ensure that the people in these roles are highly qualified. Organizations must enforce the separation of duties and maintain independent review functions (e.g., internal audit, compliance, ethics) with protected mandates, which is also a part of *role-based access control* (RBAC) (Botha & Eloff, 2001). This can be further expressed in cybersecurity governance as *five lines of accountability* (5LoA): cybersecurity control functions, the chief information security officer, internal audit, executive management, and board directors (Slapničar et al., 2023). These roles must be strongly independent, accountable, and expert. Otherwise, removing independent oversight, dismissing their need for accountability cripples, or having people in these roles who lack qualifications undermines governance integrity. This allows known vulnerabilities to persist across organizational units, especially when normal checks and balances of independence are removed. Undermining or understaffing such positions further opens the organization and executives to legal and civil liabilities. This is a problem for boards, many of which lack cybersecurity expertise (Lowry et al., 2021).

### 3.1.3. Inadequate identification of critical communication assets, risks, and governance

There was inadequate identification of critical communication assets, risks, and governance for secure communication with Signal.[ID.GV-2] The leadership involved with Signalgate did not sufficiently recognize that planning sensitive military operations via a third-party app was a critical risk, and was inconsistent with stated policy and potential legal requirements. Some officials reportedly turned to secure messaging apps like Signal during periods of administrative transition (including in the previous Biden administration), citing the lack of timely access to formal government communication infrastructure.[10] However, this use was not formally authorized for classified communication. It would not meet federal security requirements, and apparently the use by Biden administration officials was for sending secure, discreet heads-up messages—acting as notification tools rather than for full exchanges of classified information.[11] To our knowledge, there was no documented intervention to move the discussion to a secure system once sensitive information was being shared

**Lessons:** All organizations handling sensitive data (classified, confidential, private, or proprietary) should classify assets and approve secure communication tools and protocols, as well as the critical information flows involving those assets (Balboni et al., 2024). Regular risk assessments should flag when sensitive workflows occur on unsanctioned platforms so that proper solutions or controls can be instituted. If a secure tool is missing for a key process, the organization should quickly identify and address that gap before users create workarounds or use shadow IT. Otherwise, staff (or leaders) will often resort to convenient but insecure methods like shadow IT (Haag & Eckhardt, 2017; Silic & Back, 2014). Importantly, for Signalgate and organizations broadly, leaders should recognize that shadow IT is particularly common and rationalized by employees when it is used to increase their productivity and ability to communicate with each other (Silic & Back, 2014). Thus, heavy-handed controls are insufficient and may increase employee stress and frustration while doing their work. Organizations must address gaps in employees' needs and wants to work most efficiently as part of organizational security governance decisions. Unaddressed technology gaps beg for shadow IT, often from the top leaders and the best employees who are merely trying to be effective in their work. As a corollary, poorly designed and onerous systems that are not well matched to the roles and responsibilities of an organization's employees are major security risks.

#### 3.1.4. Poor inventory management of sensitive data assets requiring special handling

Because of how Signalgate took place, it appears that there was an insufficient effort to inventory and classify sensitive hardware, software, and data communications requiring special handling, not just Signal software, but also the use of personal devices for data communication that required additional safeguards.[ID.AM-1, ID.AM-2, ID.AM-4, ID.AM-5] Reports of personnel using personal devices and encrypted consumer messaging applications, such as Signal, highlight potential risks where sensitive or mission-critical communications may have occurred outside controlled environments. Regardless of employee intent, such practices undermined approved software and hardware asset tracking processes and created significant operational security and compliance risks. If left ungoverned, these workarounds bypass established inventory management protocols critical to DoD cybersecurity governance.

**Lessons:** A live inventory of hardware, systems, sensitive data, and external systems (including BYOD devices) is vital for high-trust environments (Kotenko et al., 2022; Kumar et al., 2017). Strong organizational governance of security, from top leadership through the lowest level of employees, requires a proactive approach (Posey et al., 2013; Xu et al., 2024), even an extra-role response outside one's regular duties (Hsu et al., 2015), to dynamically understand and track all of its core physical and virtual assets, including but not limited to, mobile end-user devices, network devices, and IoT (Edwards, 2024b). This is also a key part of **Zero-Trust Architecture** (Buck et al., 2021; Itodo & Ozer, 2024; Rose et al., 2020), another leading NIST security framework used extensively with NIST CSF, which requires immense business discipline and governance so that there is a clear understanding of roles, responsibilities, data, and assets, and that in real-time people are connected to the level of access they need to perform a task, and this is dynamically downgraded or removed when not needed. Such organizational transformation is challenging and requires the best of governance. However, initial empirical field research points to firms embracing Zero-Trust Architecture substantially decreasing their risk exposure and the costs of breach mediation (Adahman et al., 2022).

#### 3.1.5. Insufficient identification and classification of who should access sensitive communications

It also appears that there was an oversight in identity and access management concerning the use of Signal for sensitive communications.[**ID.AM-3**] As noted, National Security Advisor Mike Waltz[12] created a Signal chat group to coordinate discussions on military operations. Due to a contact misidentification, journalist Jeffrey Goldberg was inadvertently added to the group. His presence remained unnoticed throughout the talks, and no alerts were triggered to indicate the inclusion of an unauthorized participant from outside the organization.

**Lessons:** Organizations should enforce strict access governance for confidential discussions, maintain accurate contact lists, and use verification steps (e.g., multi-factor or positive confirmation of new participants) for group communications (Abduhari et al., 2024; Otta et al., 2023). Too many organizations have experienced disclosures from misaddressed emails or chats involving high-sensitivity communications. Verifying participants in any meeting or thread containing sensitive data should be standard practice. Further checks can be provided, such as requiring two people to approve adding an external member. Overall, email, multicast communication, video conferencing, and chat are considerable vulnerabilities in the security of virtually all organizations. Much more effort needs to be placed on securing group-based organization communications and embracing emerging solutions toward dynamic group communication security, which has been a vexing problem that researchers have struggled with for decades (e.g., Chaddoud et al., 2001), and is further complicated by 5G, VR, AR, digital twins, and emerging Metaverse environments (Chen et al., 2024; Lowry et al., 2025). An emerging proposed protocol that is promising is the RFC 9420 *messaging layer security* (MLS) protocol, designed to provide end-to-end encryption for group messaging, ensuring confidentiality, integrity, and authentication for messages exchanged within large groups.[13]

### 3.1.6. Lack of understanding of sensitive, classified, or proprietary data

There was an evident gap in understanding and enforcing protocols for handling sensitive and classified information.[**ID.RA-2**] Participants in the Signalgate incident engaged in discussions over the Signal app, sharing details such as timelines of strikes and drone operations. Although officials originally stated that no classified information was shared,[14] experts argue that the content likely contained classified elements, indicating potential misclassification and improper handling of sensitive data. This incident underscores a gap in awareness and enforcement of data classification policies. Notably, the NSA had issued a warning nearly a month prior, advising against using Signal for such communications, suggesting that this guidance was misunderstood, overlooked, or ineffectively implemented.

**Lessons:** Proper data classification and understanding of an organization's data have long been considered foundational to organizational security (Hennessy et al., 2009; Tankard, 2015). This is especially crucial nowadays because organizations need to clearly understand regulated and non-regulated data as part of security risk management (Yang et al., 2023), which is particularly challenging for global firms. However, if employees do not understand or care about these classifications or are not adequately trained on handling sensitive or private data, such technical efforts can be easily undermined. A key part of leading organizational security governance, thus, is to define and communicate data classification levels and handling rules clearly, and to motivate, collaborate with, and train employees around these (Bradford et al., 2022). Employees at all levels must be trained to recognize what information is sensitive, private, proprietary, confidential, or regulated, even if it is not labeled. Treat discussion points as sensitive when in doubt and use approved secure channels. Organizations handling trade secrets or client data should instill a mindset that "just a quick chat" can still contain crown jewels they lose control of and thus must be protected accordingly.

### 3.1.7. Ignoring or not understanding compliance requirements related to sensitive information

There was an inconsistency between handling sensitive, confidential, and classified information and the legal and procedural requirements for data protection and compliance.[ID.GV-3] Officials enabled Signal's auto-delete feature for the messages, intending to enhance confidentiality through ephemeral messaging, some after just one week, others after four, making it impossible for many to be archived before deletion.[15] This practice appears inconsistent with federal recordkeeping obligations, as the messages were not retained for oversight or archival purposes. According to NextGov, the Signal chats occurred outside government-controlled systems, bypassing backup mechanisms and data-loss prevention (DLP) controls.[16] In attempting to preserve confidentiality, the actions undermined the integrity and availability of data, two core principles of information security. Because these discussions occurred outside official systems, they bypassed existing backup protocols and DLP controls. As a result, once messages were deleted, the information was no longer available, which reduced transparency and accountability.

**Lessons:** Secure communication requires traceability, archiving, compliance auditing, and aligning security controls with legal and policy obligations (Aßmuth et al., 2024; Regueiro et al., 2021). Organizations in industries that require records retention (contracts, communications, audit logs) must ensure that encryption or auto-deletion features are still configured to meet those needs (for example, by retaining an encrypted journal of messages). "Protect" does not just mean secrecy; it also means safeguarding the reliability and traceability of data. Organizations should establish guidelines for using features like disappearing messages. Any protective measures that conflict with compliance must be addressed via policy or technical solutions, such as enterprise versions of messaging apps that archive data for compliance purposes. Organizations and researchers also need to work through thorny emerging issues related to the various traceability and archiving issues and solutions related to blockchain (Liu et al., 2022), big data (Appelbaum, 2016), remote work and devices, the Cloud (Siddesh & Rao, 2025), and AI, which are also converging with group communication solutions.

### 3.1.8. Inadequate cybersecurity governance and oversight

With the many personnel changes and the relatively quick actions against Houthi forces upon the new administration taking control, we believe an inadequate cybersecurity governance culture carried over from campaign operations into official government processes, which led to the prioritization of convenience over compliance.[ID.GV-2, ID.GV-3, ID.GV-4] This challenge would be rooted in transitioning from an informal, fast-moving campaign environment to one requiring structured oversight and mission-aligned protocols. The less formal behaviors that once enabled agility and initial formation for members of the new leadership team likely overshadowed the tighter controls required by their public-sector responsibilities, particularly in national security operations such as military planning. Though intentions were not to harm the US and its military service personnel, the desire to quickly disseminate critical information about the strike to high-level officials outweighed the guidance of cybersecurity experts and policy, resulting in unapproved communication methods and noncompliance with federal record-keeping requirements. The inability to adhere to secure communication and governance protocols during this period reflects a fundamental failure to "flip" organizational posture in alignment with the mission and regulatory expectations of the Federal office. In all fairness, however, there was a relatively tight window between the high-level officials in their respective positions and the attacks in Yemen.

**Lessons:** Organizations must establish and maintain a strong governance framework and accountability for compliance with security policies, and senior leadership must set the tone by enforcing rules even when inconvenient (AlKalbani et al., 2016; Gilbert & Gilbert, 2024). Subsequently, institutions should ensure that what is being asked of their agents is not onerous, and the level of protection truly matches the sensitivity of the task. Thus, leaders must be willing to review the tasks assigned to their subordinates to determine when it makes sense to decrease or increase the security requirements of those tasks. If everything is deemed essential and sensitive, then nothing is.

Likewise, meaningful and high-quality security policy creation, oversight, and enforcement are crucial and are increasingly tied to improved organizational security outcomes (Brown et al., 2024). Doing this poorly or without understanding and connection to employees' needs and perceptions, or in a heavy-handed, unfair manner, can backfire (Lowry et al., 2015). Regular oversight (e.g., audits or IT reviews) should catch and correct insecure practices, whether executives use personal apps for work or engineers transfer IP over personal email. An organization should empower its security officers or compliance team to intervene when protocols are skirted. Ignoring expert warnings or established policies creates governance risks that may contribute to significant security incidents.

### 3.2. Protect

In the NIST CSF, the **Protect** function focuses on implementing appropriate safeguards to ensure the delivery of critical services and mitigate the impact of potential cybersecurity events on the assets identified in the earlier CSF function. It includes activities that limit or contain the effects of a possible cybersecurity incident. Key categories within this function include access control, awareness and training, data security, information protection processes and procedures, maintenance, and protective technology. Together, these measures support a resilient security posture by reducing exposure and enhancing the organization's ability to respond effectively to cyber threats. In this function, we identified the following issues with accompanying lessons:

- Use unapproved or non-secure communication channels to transmit sensitive or protected data, violating secure communication protocols.[PR.DS-2, PR.PT-4]
- Inadequate security configuration and management of remote or mobile devices used in operational contexts, resulting in gaps in endpoint protection.[PR.AC-5, PR.PT-1]
- Insufficient access controls and authentication measures that may allow unauthorized user access to protected systems or data.[PR.AC-1, PR.AC-6]
- Failure to apply the principle of least functionality in system configurations, exposing systems to unnecessary services and increased attack surfaces.[PR.IP-1, PR.IP-3]
- Incomplete or inconsistent vetting of personnel in key security leadership roles, leading to insider risk and reduced protection of privileged functions.[PR.IP-11]
- Lack of mandatory, role-specific security awareness training and insufficient enforcement of protocol adherence, particularly for users with elevated privileges.[PR.AT-1, PR.AT-3]

#### 3.2.1. Using unapproved or insecure channels to transmit sensitive data

Signal was a known, unapproved communication channel for transmitting sensitive, confidential, and classified information, yet participants relied on it for operational discussions.[PR.IP-1, PR.PT-3] The rationale seems to have rested primarily on the app's end-to-end encryption, reflecting a limited understanding of broader security principles of data at rest. Whereas Signal encrypts content in transit, it is not an approved or vetted platform for classified communications. Its use bypasses federally mandated protections such as controlled access environments, audit logging, and hardware-level safeguards. Any compromise of a participant's device—whether through theft, malware, or unauthorized access—could expose the entire conversation. The continued use of Signal after a formal NSA warning further underscores the lack of adherence to secure communication policies and the risks of handling classified material on a platform not authorized for such use.

**Lessons:** Organizations must provide approved secure communication tools required for employee use with sensitive data and avoid relying on unvetted consumer apps (or any IT not under organizational control) that beg for security breaches (Liu et al., 2019; Onik et al., 2025). Consumer-grade tools must be strictly banned for sensitive, regulated workflows. Encryption alone is not enough if the platform or device is not hardened—the organization should ensure that confidential discussions occur only on channels that control access, encryption keys, if possible, and logging. Bring-your-own-device (BYOD) and personal devices are security risks that require special governance and meaningful job planning (Zahadat et al., 2015). Organizations must have carefully designed BYOD policies or provide carefully managed remote work devices. As part of this, organizations need to understand the relevant privacy laws and work through issues such as data and device ownership, comingled data, which providers and cloud services are involved, and so on (Ratchford et al., 2018). Firms should also regularly remind employees that seemingly convenient apps (e.g., chat, cloud drives, email) are not acceptable for confidential business interactions unless explicitly approved and secured by IT.

#### 3.2.2. Insecure Management of Remote and Mobile Devices

A core issue in the Signal case was the use of personal devices to conduct sensitive communications, which indicates limited reflection of control over remote and mobile endpoints in high-risk operations. This reduced secure encryption management and was inconsistent with best practices for endpoint protection hygiene.[**PR.AC-3**] Notably, US Special Envoy to the Middle East Steve Witkoff reportedly refrained from participating in some Signal discussions while traveling in Moscow[17], citing security concerns and choosing not to bring his phone. This example highlights an assumption that using Signal domestically was acceptable (when it was not), whereas foreign travel introduced heightened risk, which was not necessarily so. In reality, threats to mobile device security exist globally, including within the US, where digital surveillance and endpoint compromise remain persistent and exhibit well-documented risks. The incident underscores the need to enforce secure mobile device policies consistently, regardless of location.

**Lessons:** Encryption alone is insufficient without device management, threat modeling, and managing and protecting all the endpoints, of which there are many (Kamruzzaman et al., 2022). Any device accessing sensitive information should have enterprise controls: enforced strong authentication mechanisms, up-to-date security patches, remote wipe capability, and preferably mobile threat defense software. Organizations with employees handling valuable data (from R&D designs to client financials) must ensure those employees are not doing so on unsecured personal laptops or phones. If they are, the protective measures must extend to those endpoints via mobile device management or similar solutions. Recent advances in enterprise endpoint security management show that using a dynamic, real-time approach based on zero-trust architecture can increase user convenience and enterprise security (Shen & Shen, 2024).

#### 3.2.3. Weak Access and Identity Controls

There was an oversight in basic access control and identity management practices, most notably when an unauthorized individual—journalist Jeffrey Goldberg—was mistakenly added to the Signal group chat and remained unnoticed for several days.[PR.AC-1] His presence was not detected until after he left, indicating the absence of proper participant verification or monitoring. This incident reflected a gap in identity assurance: participants were added based on personal contact lists rather than a vetted directory, role-based system, or dual-authorization check. As a result, there was no systematic verification of who had access to sensitive discussions, which was inconsistent with core principles of access control and secure identity management.

**Lessons:** Organizations should assume every communication platform is a risk surface, and if identities are not verified upon invitation or login, the system is only as secure as its least verified participant (Buck et al., 2021; Itodo & Ozer, 2024). Likewise, access control must be coupled with real-time identity verification and go beyond mere authentication (Junquera-Sánchez et al., 2021). Accordingly, organizations must adopt zero trust principles: never assume internal access implies verified identity (Wu et al., 2021). There should always be multi-step verification and approval protocols before granting access to sensitive discussion threads or systems. Even internal users should be challenged to gain high-privilege access. Moreover, a key principle of access control is that access control systems must log and flag anomalies such as new additions, external contacts, or identity mismatches, especially for sensitive groups. Finally, decentralized or ad hoc access control is a red flag. Organizations should use centralized *identity and access management* (IAM) integrated with enterprise directory services (e.g., Active Directory, Okta) to enforce consistent and revocable permissions. Organizations should also automatically review access lists for sensitive systems on a scheduled basis, especially after major organizational changes or deployments.

### 3.2.4. Failure to Limit Access and Enforce Least Functionality

The Signal group did not adhere to the principle of least functionality and proper access segmentation by allowing a broad group of leaders to access and discuss information not directly relevant to their roles.[PR.PT-3] There was no clear separation of duties, and the same participants were reportedly involved in over 30 other Signal chat groups covering unrelated topics—this lack of functional boundaries created overlap in information sharing, where responsibilities and authority were blurred. Without role-based access control or information compartmentalization, sensitive discussions became unnecessarily exposed, leading to confusion over ownership, accountability, and decision-making authority.

**Lessons:** A key principle that is foundational to zero-trust architecture is the "least functionality" principle (Buck et al., 2021; Tayouri et al., 2022), which is complemented by the "least privilege" principle (Brickley & Thakur, 2021). Per **NIST SP 800-207** (**Zero Trust Architecture**) and **NIST SP 800-53 Rev. 5**, *least functionality* means that systems should be configured to provide only essential capabilities required to perform authorized functions, and no more. This means applications, systems, and services are hardened to eliminate non-essential features. Users and processes only access the minimum resources needed for their function. Default permissions, access to tools, and communication pathways are minimized by design. Zero Trust assumes breach by default, so minimizing functionality reduces the attack surface. Thus, if a user is compromised, their functional scope should not allow lateral movement or access to unrelated data/systems. If a communication tool is misused, it should be configured so that broader harm cannot be caused. When applied to organizational roles, least functionality means that employees should only be granted the tools, data, permissions, and communication channels required to perform their duties. This is part of RBAC, and increasingly, *attribute-based access control* (ABAC), which adapts access based on roles, contexts, time, device, or sensitivity (Khan, 2024).

However, least functionality is not just about "locking down apps"—it is about functionally fencing in users, devices, apps, and processes. Doing that well requires exceptional governance, leadership, and organizational alignment, or things can break down quickly. This and related Zero Trust architecture need to be infused in the organizational culture of how they work and view data, not just oversight of finance, compliance, and operations (Pigola & Meirelles, 2025). Otherwise, it can fail in situations not easily controlled through technical means, such as traditional meetings. This is where most organizations fail to apply the least functionality and separation of duties, as they do not consider meetings to be a part of the security threat landscape, when they are integral. Thus, organizations routinely make the mistake of inviting too many employees to meetings "just in case," often including people not directly involved in the issue, increasing the exposure of sensitive information, which is contrary to least-functionality principles and segregation of duties (cf. Groll et al., 2025). Without secure configuration or pre-screening, sensitive topics are discussed over Zoom (or other meeting apps). Also, research shows that the images from such meetings pose security and privacy risks (Kagan et al., 2024). There is also a lack of visibility controls: Shared screens or chat logs may inadvertently expose confidential materials. Thus, applying the least functionality to meetings would mean inviting participants essential to the task or decision; using platforms that enforce authentication and access controls; restricting what can be shared, recorded, or forwarded; and ensuring clear data classification protocols are followed during the meeting.

#### 3.2.5. Inaccurate Vetting of Security Leadership

During the transition to the new administration, several key national security officials entered their roles under expedited conditions, consistent with exceptions granted to presidential appointees. On January 20, 2025, President Trump issued a memorandum directing the White House counsel to grant temporary Top Secret / Sensitive Compartmented Information (TS/SCI) clearances to designated personnel—bypassing the usual depth of vetting.[18] These interim clearances were valid for up to six months while background investigations were pending. Although these exceptions are legally permissible, they also mean that some participants—despite being involved in high-level discussions—had not completed traditional background investigations as outlined in Executive Order 10450 or through the SF-86 process.[PR.IP-11] Public reporting has noted that some appointees had prior personal or professional risks that would typically prompt further review under standard vetting procedures, with Hegseth being the primary person of concern, with reports of the administration's transition team being "blindsided" by the new details of allegations they were not aware of. [19]At the same time, Inspectors General responsible for oversight were dismissed or their positions left unfilled, and Congressional confirmation processes drew criticism for insufficient scrutiny. Together, these factors contributed to a fragmented personnel security posture during a sensitive period of national security decision-making.

**Lessons:** Organizations must include cybersecurity as a core part of their human resources practices (e.g., de-provisioning, personnel screening, recruiting, onboarding, and off-boarding) such that **PR.IP-11** emphasizes that a cybersecurity and privacy workforce should be recruited and screened to verify that individuals meet established security criteria (Alamoudi, 2022; Llorens, 2023; Nica et al., 2024). This control ensures that individuals entrusted with sensitive systems, data, or decision-making authority are reliably vetted, cleared, and capable before they are granted such access. No one should be exempt, especially an organization's leaders. High-level hires must undergo formal security vetting, including background checks, ethics disclosures, and insider threat assessments.

Moreover, screening must include suitability, not just qualifications. As noted, in some cases, appointees such as Hegseth had personal or professional vulnerabilities that would typically warrant additional scrutiny in sensitive roles, suggesting a need for more detailed review. In selecting people for positions, HR and security must coordinate to assess risk-based attributes such as financial stress, prior misconduct, or susceptibility to blackmail, not just résumé strength or political alignment.

Finally, organizations must resist pressures to expedite the onboarding of senior leaders without due diligence. Leadership roles often have the greatest access and least oversight, making screening more, not less, critical. Rushed onboarding short-circuits hiring controls. In times of transition (e.g., mergers, reorganizations, elections), security and HR must collaborate on fast-track vetting pipelines that still meet baseline standards.

#### 3.2.6. Insufficient Security Training or Enforcement of Protocol, Especially for Privileged Users

Due to the accelerated transition and confirmation process, several key participants did not receive adequate security training to handle confidential and classified information, particularly privileged users with elevated access.[PR.AT-2] Notably, no one in the Signal group raised objections to using an unapproved, non-government platform for sensitive communications, suggesting either a lack of awareness of applicable policies or a culture that did not prioritize adherence to established security protocols, or too many conversations on the same platform, muddying the demarcation between non-sensitive and sensitive topics. This situation reflects a gap in both initial and ongoing training, as well as weak enforcement mechanisms. Continuing campaign-era communication habits, including using Signal, even after assuming official roles, points to insufficient onboarding regarding cybersecurity expectations. Separately, media reports later indicated that former official Hegseth was reported to have shared confidential military operational details via Signal with individuals who were not cleared to receive such information, including his wife and personal attorney.[20] These incidents underscore the critical need for role-based training, more vigorous policy enforcement, and cultural alignment with federal cybersecurity standards, especially for those in senior or sensitive positions.

**Lessons:** No employee, regardless of title, should be exempt from initial and recurring *security education, training,* and *awareness* (SETA) programs that are carefully designed and tied to their role and risk exposure (Hu et al., 2021; Silic & Lowry, 2020). This is especially true of top leadership, who set the example for the entire organization. Organizational security leadership must continuously train and remind all personnel, including top executives, of security policies and why they matter. Staff hired from outside the existing organization (e.g., new hires, campaign staff, contractors) should undergo onboarding that separates personal technology habits from approved corporate practices. It is not enough to issue a policy; organizations must follow up by monitoring compliance and correcting violations, and fostering a security awareness culture (Khando et al., 2021; Tsohou et al., 2015) and security compliance culture (Alshaikh, 2020; Sharma & Aparicio, 2022) from a collective understanding of protecting the organization and its stakeholders, not blind obedience. A culture of security needs to be enforced and supported where people feel comfortable speaking up against inappropriate or unsafe practices, even when suggested by leadership.

### 3.3. Detect

In NIST CSF, the **Detect** function focuses on developing and implementing the appropriate activities to identify the occurrence of a cybersecurity event promptly. This includes continuously monitoring information systems and assets, detecting anomalies and potential threats, and understanding the normal operation baseline to identify deviations. Key categories under this function include anomalies and events, continuous security monitoring, and detection processes. Together, these enable organizations to promptly recognize indicators of compromise and initiate appropriate response actions, ensuring that threats are detected early enough to minimize impact. These led to the following issues and lessons we have identified with Signalgate:

- Inadequate implementation of real-time monitoring and alerting for unauthorized access or potential data exfiltration.
- Limited visibility and monitoring of unauthorized or unmanaged IT assets and communication channels, often called "shadow IT".
- Failure to detect deviations from user behavior baselines, including potential policy violations or insider threat indicators.
- Insufficient integration and use of external cybersecurity threat intelligence sources to support detection and analysis efforts.

### 3.3.1. Lack of real-time detection of unauthorized access or data leakage

Because the Signalgate group operated outside of established federal communication protocols and security technologies, they operated without real-time detection capabilities that could have identified unauthorized access or potential data leakage.[DE.AE-2] The accidental inclusion of journalist Jeffrey Goldberg was one example of this oversight. The inability to determine whether external actors may have silently accessed these communications raised additional concerns. Given the use of Signal—a consumer-grade app not approved for classified or sensitive government discussions—experts have raised credible concerns about the potential for broader compromise.[21] Importantly, at the end of February 2025 (a month before Signalgate), the NSA had issued a formal directive against using Signal for such activities.[22] Reports suggest that this group and numerous other Signal chats involving overlapping personnel discussed various national security and policy matters. Security analysts have warned that without proper monitoring and audit logging, these communications could have exposed confidential deliberations on issues such as military operations, international diplomacy, trade policy, and domestic legal strategy. Although no confirmed breaches have been disclosed, the lack of visibility created uncertainty about potential security risks.

**Lessons:** Organizations must flag and follow up with unauthorized access warnings in real time (e.g., logins, participation), which involves integrated monitoring and alerting of sensitive resources and communications (Bhuyan et al., 2017; Ghafir et al., 2016; Kebande et al., 2021). For example, using a collaboration tool, enable features that notify the administrators of new participants or anomalous access. Organizations should generally have intrusion-detection or anomaly-detection systems to pick up unusual events (García-Teodoro et al., 2009; Yang et al., 2022) (e.g., large data transfers or an unknown device accessing confidential data). Even policies as simple as out-of-band verification ("call to confirm when adding a new member to a sensitive thread") are a form of detection control. The goal is to catch misrouting or unauthorized access immediately, rather than days or weeks later.

### 3.3.2. Lack of visibility into "shadow IT" channels

There was a significant gap in visibility into unauthorized, unmanaged communication channels, via "shadow IT", particularly regarding individuals using Signal on personal devices with privileged access to secure government environments.[**DE.CM-7**] Despite operating within highly sensitive organizations such as the DoD, FBI, CIA, and NSA, some participants reportedly used Signal on personal smartphones[23] without centralized oversight. In this context, Signal functioned as an unsanctioned communication platform, bypassing approved enterprise systems. As a result, standard security controls such as *security information and event management* (SIEM) logging and DLP monitoring were not in place. Security teams lacked visibility into message content, access patterns, or anomalies that might indicate compromise. Experts in cybersecurity have consistently warned that unmonitored applications with a lack of visibility—especially in classified environments—can present heightened risks, including inadvertent disclosure or targeted exploitation by adversaries (Afzal et al., 2021; Haag & Eckhardt, 2024; Walterbusch et al., 2017).[24] Moreover, the use of encrypted apps such as Signal has been shown to hinder forensic investigations (Afzal et al., 2021). Notably, in response to internal restrictions on mobile device use within secure facilities, Hegseth allegedly directed IT personnel to install a secondary workstation with external internet access (aka a "dirty line") inside the Pentagon, enabling continued use of Signal.[25] This workaround highlights the tension between usability and policy enforcement, and it underscores the importance of ensuring that all communication platforms used in high-security contexts are appropriately vetted, monitored, and compliant with established cybersecurity protocols.

**Lessons:** Organizations need to maintain visibility on where sensitive data is going, what applications are using the data, and if the applications are approved for the given use context for specific roles (e.g., are email and Zoom being used as intended?) (Alqahtani et al., 2024; Brant et al., 2021). Organizations should inventory and monitor approved communication paths, and actively hunt for unapproved ones to ferret out shadow IT (Haag & Eckhardt, 2024; Silic & Back, 2014) and unapproved workarounds (Alter, 2014; Besnard & Arief, 2004), even if for legitimate or well-intended purposes. This might involve periodic audits of employee devices or network traffic using unapproved apps. *For instance, cloud access security broker (CASB) solutions* can identify usage of unauthorized cloud services (Ahmad et al., 2022). The key lesson is that you cannot detect what you cannot see: if employees are discussing proprietary information on an off-channel (e.g., WhatsApp, personal email, Signal), the organization needs a strategy (policy + technical) to either bring that activity into the light or shut it down.

### 3.3.3. Failure to detect policy violations and unusual insider behavior

There was an apparent gap in monitoring and detecting off-policy communication behaviors or potential insider risk activities, and a limited ability to trace or audit those interactions retrospectively.[DE.CM-1] Reports suggest that government officials frequently created Signal chat groups, often named after ongoing policy crises or operations, such as the "Houthi PC small group." This naming pattern may indicate a broader trend of using unapproved communication tools for sensitive discussions, yet no internal detection or audit mechanisms flagged this behavior in real time. Additionally, using Signal's disappearing message feature severely limits retrospective review. If routine compliance checks were in place, they lacked visibility into these channels, leaving oversight bodies without meaningful audit trails. Ultimately, rather than internal monitoring systems, an external watchdog investigation and legal action led to attempts to recover or account for the missing communications. This case highlights the risks posed by a lack of endpoint monitoring, inadequate policy enforcement, and insufficient insider threat detection capabilities, especially in high-trust, high-clearance environments.

**Lessons:** Technologies and technical controls cannot catch all security threats; thus, security policies and how they are understood and enforced are often the last, and sometimes most important, organizational security defense (AlKalbani et al., 2016; Hu et al., 2012). Regular compliance reviews should flag if, for example, there are no emails or meeting minutes about a major project, which may indicate the conversation has moved to an unmonitored medium. Unusual behavior, like a sudden absence of communication on official platforms or the appearance of a new tool, should be investigated as a potential security gap. There must also be whistle-blowing protocols, systems, policies, and support mechanisms to report serious violations without reprisals (Aldrich & Moran, 2019; Lowry et al., 2013). It is difficult for us to imagine that a Pentagon IT staffer would not have known that installing a dirty Internet line as a workaround for confidential and secure communication on Signal was an extreme security policy violation. Whether someone spoke up about this violation is unknown; however, such a situation is rife with issues related to power imbalances, thereby making available confidential whistle-blowing systems that are much more important.

### 3.3.4. Not leveraging cyberthreat intelligence (CTI)

A key issue in this case was the lack of *cyber threat intelligence* (CTI) initiatives to inform detection and response protocols, particularly regarding using non-sanctioned communication platforms like Signal.[DE.CM-8] Public reporting indicates that the NSA—a primary CTI source for the DoD, CIA, FBI, and other national security agencies—had explicitly warned that foreign state actors were actively targeting Signal due to known vulnerabilities. Despite these warnings, participants continued to use Signal for discussions involving sensitive national security matters.

The US government has invested extensively in CTI infrastructure, and the NSA's advisories are intended to support proactive, organization-wide cyber risk management. From a governance standpoint, it is notable that senior officials in high-security agencies did not act on this advisory, with no immediate pivot in communication protocols, policy reinforcement, or retroactive monitoring of platform usage. In high-trust environments handling sensitive or classified information, an NSA directive regarding application security should prompt a comprehensive review of communication tools, enforcement of approved platform usage, and incident response procedures. The absence of such coordinated action indicates challenges in CTI integration and operational enforcement at the leadership level.

**Lessons:** Mature security operations continuously integrate real-time intelligence about attackers' tools and techniques to catch them quickly if they appear in the organization's environment. Consequently, building on *technical threat intelligence* (TTI) (Tounsi & Rais, 2018), leading organizations fully embrace CTI efforts as one of the few ways to not always be in a defensive posture, but to instead aggressively collaborate with other leading organizations, governments, and real-time intelligence to prepare for likely future events and to even go on offense with counter-intelligence and counter-attacks (Ainslie et al., 2023; Shin & Lowry, 2020). Such efforts must focus on an organization's highest-value assets. Moreover, when trusted partners and government entities warn that a specific attack vector is on the rise (e.g., phishing of executives' personal messaging accounts), organizations must rapidly respond by enhancing monitoring, vulnerability scanning, or protections in that area, and informing all parties using such assets. This could mean instituting phishing-resistant *multi-factor authentication* (MFA), *endpoint detection and response* (EDR) on devices that handle sensitive data (Chen et al., 2023), or alerting users to be on high alert for specific tactics. The correct security posture is to take the warning seriously and assume that this means there is already a likely bad-actor breach, and to immediately operate accordingly to find the breaches and shut down the vulnerabilities, and to leave no stone unturned until the organization can confirm there were no breaches, and if there were breaches, to execute the proper pre-planned response plan.

### 3.4. Respond

In the NIST CSF, the **Respond** function emphasizes the need for well-defined procedures to manage and mitigate cybersecurity incidents effectively. This includes having a structured incident response plan, enabling coordinated communication with stakeholders, ensuring prompt mitigation, and continuously improving based on lessons learned. In this case, several weaknesses were observed in how the organization responded to the cybersecurity event, reflecting gaps across multiple categories within the Respond function:

- Inconsistent or poorly coordinated incident response.
- Delayed containment and inadequate initial mitigation.
- Insufficient incident communication and regulatory reporting.

#### 3.4.1. Inconsistent or Poorly Coordinated Incident Response

There is no public evidence that the leadership involved in the Signalgate incident followed an established or tested incident response plan.[RS.RP-1] The reaction to the incident appeared disorganized and characterized by public reassurances rather than structured containment and root-cause analysis. Several officials publicly downplayed the significance of the breach. For example, the Defense Secretary responded with personal criticisms of the reporting rather than addressing the substance of the alleged security lapses, stating that "nobody was texting war plans." Similarly, media reports suggest the President's initial concern focused more on how a journalist's contact ended up in a staff member's phone than on the potential security implications of unvetted access.

These responses suggest that no formal response playbook was activated to manage the incident or guide coordinated communication and mitigation efforts. Instead, the administration's response focused more on managing public messaging than immediate remediation and analysis. The absence of a visible framework for classifying the event, assigning response roles, containing further exposure, or issuing public guidance indicates a gap in applying fundamental incident response protocols.

**Lessons:** All organizations need defined, rehearsed incident response plans and protocols, including an understanding of how incidents are named and classified, the stakeholders involved, and how to communicate about specific incidents (Alfred et al., 2025; Death, 2017; Thomas et al., 2022; Thompson, 2018). Organizations should treat incident response as a capability that needs to be developed, focusing on a collaborative approach with organizational leadership (Ahmad et al., 2012; Ahmad et al., 2021; Mitropoulos et al., 2006). In a corporate setting, if an executive's emails leak, for example, the response should be coordinated with the security incident team and follow a known, practice protocol for the response, not left to the executive alone to respond independently without guidance. Table-top exercises with executives can help prepare them to respond calmly and effectively in a coordinated fashion with security leadership. A prepared organization will address the breach itself (what happened, what do we do next) before trying to manage public perception.

#### 3.4.2. Delayed Containment and Inadequate Initial Mitigation

Once the incident was reported, there was no publicly available evidence that officials initiated an immediate technical response to contain the incident, such as revoking access, notifying participants, or migrating communications to an approved secure channel.[RS.RP-1, RS.MI-1] Signal use reportedly continued across multiple chat groups, including during and after the breach window, while the auto-delete features remained enabled. This may have impeded the ability to preserve forensic evidence, complicating post-incident investigation and limiting visibility into the extent of exposure.

From a cybersecurity perspective, not promptly disabling a potentially compromised communication channel could have allowed any ongoing incident—if present—to continue undetected. In scenarios involving surveillance or eavesdropping, delays in containment actions increase the risk of further data leakage. No publicly disclosed actions suggest a coordinated effort to isolate the incident across all related chat groups or enforce temporary restrictions while an investigation was conducted. This underscores possible shortcomings in incident containment planning and execution consistent with best practices under the NIST CSF Respond function.

**Lessons:** One of the most pressing issues in proper security incident response is rapid isolation and containment (Ahmad et al., 2012; Alfred et al., 2025; Mitropoulos et al., 2006). As soon as an incident is identified, steps like disabling compromised accounts, removing unauthorized users, or taking affected systems offline should occur to prevent further damage. In many data breaches, the speed of response can mean the difference between a minor incident and a major loss, and has been directly tied to greater costs when not immediately resolved. Organizations should pre-plan containment strategies for various scenarios: e.g., what if a confidential group chat or email thread gets an unintended recipient? Have a procedure (such as immediately instructing participants not to share more until IT evaluates, confirming the recipient's deletion of messages, etc.). Swift action can limit exposure and demonstrate control over the situation.

#### 3.4.3. Insufficient Incident Communication and Regulatory Reporting

Throughout the handling of the incident, communication efforts with key stakeholders were inconsistent and were limited in transparency, falling short of controls and best practices involving internal coordination.CO-1] stakeholder communications,[RS.CO-2] and information sharing with external stakeholders.[RS.CO-4] The administration's response showed limited collaboration and delayed disclosures to oversight entities, reducing confidence in the incident handling process. Rather than following a coordinated communication strategy, public statements appeared disjointed and reactive, sometimes minimizing their significance or declaring them resolved before investigations were complete. This approach contributed to continued scrutiny and stakeholder dissatisfaction. Key oversight bodies, including Congress and the judiciary, were not proactively engaged. It took court orders to prompt the preservation of message logs, many of which had already been auto-deleted when investigators were granted access. Congress was required to initiate formal inquiries to obtain information typically provided through standard incident response protocols.

The communication breakdown extended beyond internal governance. Reports indicated concern from international allies whose operational details may have been exposed. This would resemble a data breach disclosure failure in a private-sector context, where clients, regulators, and partners are not fully informed, risking reputational and operational harm. The response lacked a structured communication plan with affected parties, including record-keeping bodies, allied governments, and the public. This absence of incident communication and accountability leadership created further confusion and hindered resolution efforts.

**Lessons:** A clear chain-of-command for crisis escalation is vital in time-sensitive scenarios, and an effective incident response plan includes immediate steps for communication. Importantly, it calls for fact-based updates rather than deflection, with a pre-planned understanding of who is accountable for communicating to stakeholders, including their remediation plans (Gwebu et al., 2018; Knight & Nurse, 2020; Thomas et al., 2022). Organizations must be better prepared to handle crisis communication principles that consider the correct messengers, effective and accurate messaging, and mindfulness of the key stakeholders who should be communicating with them. (Coombs & Holladay, 2010; Liu et al., 2011), which is especially crucial in contexts of high uncertainty that security incidents create (Knight & Nurse, 2020). Internal confusion can undermine trust with regulators, clients, and investors. Communication playbooks are essential for breach containment and brand protection. Effective incident response includes clear communication plans. Organizations must identify who needs to be informed (e.g., board members, executive leadership, legal counsel, customers, partners, regulators) and in what timeframe. Failing to communicate proactively often backfires by eroding stakeholder trust and inviting external enforcement. It is better to proactively disclose what is known and what steps are being taken, even if the whole picture is unclear. In many industries, there are also legal obligations for breach notification. Thus, failing to communicate can turn a technical issue into a reputational crisis. Organizations should designate spokespersons and internal liaisons for incidents and have pre-drafted guidelines for messaging to various audiences (without revealing sensitive details that could further compromise security).

### 3.5. Recover

The **Recover** function in the NIST CSF focuses on restoring capabilities and services following a cybersecurity incident while reinforcing organizational resilience. This includes executing recovery plans, integrating lessons learned, and rebuilding trust through coordinated communications. Effective recovery helps minimize the long-term impact of an incident and ensures preparedness for future disruptions. The following recovery challenges were identified in this case:

- Lack of transparent leadership accountability

- Inability to recover or reconstruct critical communication data
- Absence of a formal remediation or recovery plan
- Failure to institutionalize lessons learned
- Limited action to rebuild public and institutional trust

### 3.5.1. Lack of Transparent Leadership Accountability

Throughout the Signalgate incident, public accountability for leadership decisions was limited, which made it more difficult to rebuild trust—an essential component of recovery.[RC.CO-1, RC.CO-2] Despite the seriousness of the situation, immediate accountability actions were limited, and the communication response was criticized by observers as insufficiently clear, raising concerns about the effectiveness of the crisis management approach.

Notably, the National Security Adviser who coordinated the Signal discussions retained his position, with administration officials reportedly defending using Signal as a necessity given the lack of a readily available alternative. However, this rationale—absent accompanying corrective actions— risked creating the impression that established security protocols are flexible or optional, especially for senior leadership. This perception can be problematic in national security organizations like the DoD, NSA, CIA, and FBI, where all personnel expect strict adherence to chain-of-command structures and information security rules. When accountability is unclear after a significant breach, it may create the perception that cybersecurity compliance is negotiable, which can weaken security culture and erode standards if employees believe violations carry no consequences

**Lessons:** Organizations that are inconsistent in embracing accountability for internal security incidents and violations are more likely to suffer from decreased staff morale, client trust, and general compliance, and are more likely to have future breaches than highly accountable organizations (Chen et al., 2021; Oluka, 2023). After a security incident, there must be follow-through: analysis of root causes, assignment of responsibility, and implementation of fixes (Ahmad et al., 2021; Sahu, 2024). Accountability does not always mean firing someone, but it should mean that the organization addresses any negligence or ignorance through appropriate measures (could be re-training, policy change, or disciplinary action if warranted). The key is to avoid a culture of complacency. Security and governance professionals should document the incident in an after-action report and track the remediation steps. Not only does this resolve the specific issue, but it also signals to all employees and stakeholders that the organization is serious about its security protocols and will act promptly and appropriately when they are violated.

#### 3.5.2. Inability to Recover or Reconstruct Critical Communication Data

Due to the use of communication platforms with ephemeral messaging features—specifically Signal's disappearing message function—there was a considerable constraint in recovering critical data and records after the incident, undermining forensic investigation and organizational learning.[RC.IM-1] When oversight bodies attempted to retrieve chat content, only metadata such as group names and participant lists were available; the messages had already been deleted. This resulted in losing essential communications related to key decisions and operational actions.

In secure communications governance, this represents a gap in data retention and incident documentation, both essential for accountability and for analyzing root causes. In a corporate analogy, it would be akin to losing all records of a major contract negotiation conducted over an unlogged, self-deleting platform, leaving the organization vulnerable in case of legal disputes, compliance inquiries, or internal reviews. Given that Signalgate only involved a single chat group of the over 30 Signal chats that were not revealed, the true scope of unrecoverable data and potential wider compromise is unknown. Any security review is incomplete without clear logs or retained transcripts, and critical lessons that could improve future protocols are missed.

**Lessons:** Organizations need to incorporate high levels of data resilience and recovery auditing into their security strategies, in which real-time, reliable, and accurate backups and record-keeping are part of recovery (Edwards, 2024a; Liu. et al., 2021; Syed, 2024). If communications or decisions are important enough to affect business or security, ensure they are not ephemeral and beyond retrieval, or there is a mismatch between strategy and resiliency. Suppose there are appropriate times for using a communication tool that does not automatically archive. In that case, they need to establish a process to capture essential information (even if it manually records meeting minutes for an encrypted call, for example). After an incident, having logs and data to analyze is crucial for learning what went wrong. For organizations with legal or compliance mandates, it is also vital to investigate, report, and avoid further civil and legal consequences from negligent data handling.

#### 3.5.3. Absence of a Formal Remediation or Recovery Plan

The combination of disappearing messages, limited public acknowledgment of missteps, and delayed or constrained oversight significantly hindered the ability to conduct thorough post-incident remediation planning and to communicate those plans transparently to stakeholders.[RC.IM-1; RC.CO-3] As noted, the administration described the incident as relatively innocuous and did not provide a formal path for reviewing or improving incident handling processes.

A contributing factor was the disruption of typical oversight mechanisms. Several IGs responsible for ensuring agency compliance with laws and internal controls—including cybersecurity and records management—were dismissed or faced political pressure during this period. Although IGs do not have direct authority over the President or Vice President, their role in reviewing agency-level incidents (e.g., at the DoD or State Department) can be crucial. The weakened IG framework reduced the likelihood of an independent inquiry into possible violations.

Additionally, key communications were lost, and legally required recordkeeping practices were not followed due to using encrypted, auto-deleting messaging apps like Signal. This made it more difficult for internal and external reviewers to determine what had occurred or extract lessons for future response planning. The lack of a coordinated, transparent recovery effort ultimately limited opportunities for institutional learning and reduced stakeholder trust.

**Lessons:** Organizations must communicate clearly and promptly with clients, partners, oversight bodies, and other stakeholders following a cybersecurity breach and treat this as a form of crisis communication and management (Knight & Nurse, 2020; Ruohonen et al., 2024; Sapriel, 2021). In the United States, publicly traded companies are legally required to report material cybersecurity incidents within four business days under new SEC disclosure regulations issued in 2023. This legal mandate reflects a broader expectation that transparency is not optional—it is essential for maintaining public trust and ensuring regulatory compliance.

Beyond meeting external disclosure requirements, effective recovery also depends on an honest and thorough internal assessment. To the best of their ability, organizations must determine how the breach occurred, identify gaps in oversight or governance, and assess the failures in systems, policies, or individual behavior that allowed the incident to happen. This process should also include concrete remediation steps and a clear plan for improving resilience. Significantly, these actions should not be limited to internal audiences. Stakeholders— including clients, partners, and regulators—have a right to know that a breach occurred, how the organization responds, and what changes are being made to prevent recurrence.

When organizations minimize, withhold, or delay disclosure of an incident, they risk legal consequences and undermine their credibility. Communication gaps, vague public statements, failure to take responsibility, or unclear accountability can foster mistrust and frustration among key stakeholders. The recovery process, therefore, must go beyond restoring systems or operations—it must also prioritize restoring confidence through honest engagement, clear messaging, and a demonstrated commitment to accountability. Anything less risks compounding the damage of the original incident.

### 3.5.4. Failure to Institutionalize Lessons Learned

A significant obstacle to institutionalizing lessons learned from the Signalgate incident was the absence of clear accountability and transparency following the breach.[RC.IM-1; RC.IM-2] Key actions and decisions were minimized in public discussions, and there was limited acknowledgment of governance or procedural failures. One underlying contributing factor was the lack of an official, real-time, secure messaging platform that could be used for urgent inter-agency communication. Whereas SCIFs (Sensitive Compartmented Information Facilities) remain the "gold standard" for classified discussions, their limited accessibility makes them impractical for dynamic coordination in fast-moving discussion scenarios involving multi-agency discussions or large working groups concerning the Administration. SCIFs are a classic example of a design gap, illustrating reliance on legacy solutions that do not fully align with modern communication needs, which can make daily operations cumbersome and encourage workarounds such as shadow IT (Haag & Eckhardt, 2024; Silic & Back, 2014)..

After the incident, this structural gap persisted, as no immediate secure alternative was deployed. Without a designated platform, personnel may default to less secure tools out of necessity, particularly under time pressure. This highlights a broader issue in recovery planning: failing to address the original process deficiencies that contributed to the incident in the first place. Without a defined and timely "Plan B," organizations remain vulnerable to similar behaviors or technical workarounds, even after a breach.

This pattern is not unique to the government. In private-sector examples, companies that experience breaches due to outdated infrastructure—such as legacy VPNs—often struggle to implement new secure systems quickly, exposing them to continued risk. Addressing recovery effectively means improving technical safeguards and remediating the operational and behavioral drivers that create incentives for noncompliance. Again, we believe the noncompliance evident in Signalgate was non-malicious in nature. Regardless, a more robust and transparent post-incident strategy would have included policy reinforcement and investment in viable, secure alternatives for real-time communication.

**Lessons:** Open and honest organizational learning and change are fundamental to effective organizational security incident response (Ahmad et al., 2012; Ahmad et al., 2021; Bitzer et al., 2023; Mehrizi et al., 2022). Learning after an incident as an organization may be one of its most crucial cybersecurity decisions because it strengthens operational resilience by closing process and technology gaps identified by incidents, and improves an organization's security and learning culture (Burns, 2019; Mehrizi et al., 2022; Patterson et al., 2023). Recovery is not just about getting systems back online but about restoring operations to a more secure state and strengthening an organization's security posture. If a breach happened because an approved tool did not exist (forcing users to improvise), a key recovery step is to deploy a proper tool as soon as possible. In practice, this means budgeting and planning for secure infrastructure before an incident, if possible, but accelerating those improvements afterward. Until the new solution is in place, interim guidance—such as stricter rules or temporary workarounds—should be given to avoid falling back into the same risky behavior.

### 3.5.5. Limited Action to Rebuild Public and Institutional Trust

There was no apparent, coordinated effort to rebuild trust or organizational credibility following the breach. Instead, some post-incident personnel decisions and public statements that appeared were perceived as reactive and likely did not help restore confidence.[RC.IM-1, RC.CO-2] Reports of high-level dismissals—including close aides and senior security officials—were not accompanied by transparent explanations or a clear roadmap for restoring stakeholder confidence. Rather than reinforcing accountability and competence, these actions created uncertainty about internal cohesion and governance.

The Signalgate incident significantly eroded public trust in the government's ability to manage sensitive information securely. The perception of opacity and inconsistent messaging, alongside headlines describing the situation as "undermining transparency and accountability," amplified concern among both domestic and international stakeholders. In any organization, a high-profile data breach or operational failure requires more than technical remediation; it necessitates deliberate efforts to restore confidence among employees, partners, clients, and oversight bodies.

Failure to acknowledge mistakes or articulate a forward-looking recovery plan may worsen reputational damage. Internally, it may lead to diminished morale or a breakdown in confidence in leadership. Externally, stakeholders may question whether future lapses will be addressed with the seriousness required. The Signalgate response highlights the importance of pairing post-incident recovery actions with clear, public-facing communication strategies that demonstrate responsibility, outline reforms, and commit to rebuilding trust over time. Without such efforts, organizations risk compounding the impact of the original breach by allowing reputational damage to persist unaddressed.

**Lessons:** An organization's security incident recovery is not complete until trust of all stakeholders is restored; thus, minimizing or overlooking a major trust violation is never an effective plan; instead, it exacerbates the trust damage and fosters lasting stakeholder distrust (Gillespie et al., 2014; Knight & Nurse, 2020; Sapriel, 2021). Trust repair after any incident is crucial (Choi & Nazareth, 2014; Gwebu et al., 2018), and when it is not conducted effectively (or ignored entirely), it can undermine insider morale to the point where future insider incidents are more likely (Lowry et al., 2015). External trust damage and repair are also crucial because otherwise, it can lead to permanent market and brand damage if not handled effectively (Choi & Nazareth, 2014; Gwebu et al., 2018). This involves transparent communication about what happened, what is being done to fix it, and visible changes to prevent a recurrence. For a business, that could mean outreach to key clients to assure them their data will be handled more securely, or publishing a post-mortem and mitigation plan if appropriate. It may also involve supporting those impacted (e.g., credit monitoring in a consumer data breach). Internally, leadership should reaffirm its commitment to security and possibly bring external auditors or consultants to validate improvements. The lesson is that technical fixes alone won't heal reputational damage; proactive and sincere stakeholder engagement is needed.

### 3.6. Govern

The sixth and final core function of the CSF Framework, **Govern**, was added in 2024 to strengthen organizational alignment and accountability around cybersecurity risk. This reflects the growing recognition that effective cybersecurity is not just a technical issue but a governance and leadership responsibility. Accordingly, the "govern" function ensures that an organization's cybersecurity strategy, policies, and risk management practices are aligned with its mission, legal and regulatory obligations, and broader enterprise risk posture. It addresses the organizational context, roles, responsibilities, and oversight necessary to support all other cybersecurity efforts. Accordingly, the governance function focuses on the following:

- Establishing organizational context: Understanding the mission, stakeholder expectations, legal, regulatory, and contractual requirements influencing cybersecurity decisions.
- Defining risk management strategy: Setting priorities, constraints, risk tolerances, and assumptions to support
- Assigning roles and responsibilities: Clarifying cybersecurity roles, responsibilities, and authorities to foster accountability and continuous improvement.
- Developing policies and procedures: Creating and maintaining policies that reflect organizational objectives and compliance requirements.
- Ensuring oversight: Monitoring and reviewing cybersecurity strategies to adapt to evolving risks and organizational changes.7
- Managing supply chain risks: Identifying and mitigating third-party suppliers and partners' risks.

These categories collectively ensure cybersecurity considerations are embedded into the organization's governance structures and decision-making processes. The governance function is the cornerstone for the five traditional CSF functions: (1) **Identify**: Governance establishes the context and criteria for identifying assets, systems, and risks. (2) **Protect**: The policies and strategies defined in the governance function guide the implementation of protective measures. (3) **Detect**: Governance establishes the parameters for monitoring and detecting cybersecurity events. (4) **Respond**: Governance helps establish the roles and responsibilities to ensure effective incident responses align with organizational policies and strategies. (5) **Recover**: Governance frameworks support recovery planning and incorporate lessons learned into future policies and strategies. By placing "govern" at the center of the CSF wheel, NIST highlights its integral role in shaping and supporting the cybersecurity risk management lifecycle, including policies, oversight, and organizational alignment, making the other five functions actionable and effective.

In applying the "Govern" function of the NIST CSF 2.0 to Signalgate, we find several governance deficiencies that likely contributed to the incident. First, we believe there was a misalignment between agency missions (e.g., CIA, DoD, FBI) and how those missions should inform cybersecurity risk decisions, including protecting core outcomes, capabilities, and services.[GV.OC-01; GV.OC-05] Instead of being guided by mission-critical risk management practices, many decisions were shaped more by short-term political objectives than long-term institutional cybersecurity priorities.

Second, the interests and expectations of key internal and external stakeholders—such as agency personnel, oversight bodies (e.g., IGs and Congress), US allies, and the public—were not sufficiently considered in cybersecurity decisions or communication strategies, indicating a breakdown in stakeholder alignment and governance transparency.[GV.OC-02; GV.OC-04]

Finally, actions taken throughout the incident suggest limited regard for established legal, regulatory, and oversight responsibilities, including records retention, privacy, and coordination with institutional oversight mechanisms.[GV.OC-03] These gaps illustrate the importance of embedding governance principles at the core of cybersecurity practices to ensure resilience, accountability, and public trust.

**Lessons:** Effective top-down security governance, where the correct tone is set at the highest levels of leadership, and where security is treated as integral rather than secondary, is crucial to any organization that seeks cybersecurity resilience (Aksoy, 2025; Armour, 2017; Gilbert & Gilbert, 2024). The "Govern" function is a particularly timely addition to NIST CSF. It highlights that an analysis of Signalgate suggests the incident might have been avoided had leadership and governance priorities been more aligned with security needs. This could be argued as a case where governance shortcomings played a central role (in fact, in several other major security cases, the organization did not have a CISO in place, or executive leadership countermanded the CISO). The US Federal organizations involved did not lack cybersecurity training, oversight, protocols, technologies, or resources. Regarding cybersecurity, the US DoD, FBI, CIA, and NSA are among the most capable globally, with substantial budgets and expert personnel. However, massive resources can never overcome unclear priorities, governance gaps, or short-term decision-making that treat security as a secondary concern. The Administration's governance "tone at the top" (Johnson & Goetz, 2007) was described as weak and undermined the organizational security culture (da Veiga et al., 2020; Van Niekerk & Von Solms, 2010) at multiple US agencies that took decades to build.

Fundamental to the governance challenges observed in this case was a lack of clearly defined accountability, oversight, and leadership responsibility. Across multiple dimensions, the incident revealed ambiguity in ownership of cybersecurity decisions and unclear chains of authority, which was exacerbated by having multiple leaders from various agencies involved. In traditional organizations, effective security governance begins with strategic oversight by the board of directors, is implemented through an empowered and independent CISO, and is supported by the CEO and other C-level executives to embed a security-conscious culture throughout the organization (AlGhamdi et al., 2020; Johnson & Goetz, 2007; Johnston & Hale, 2009; Posthumus & von Solms, 2004). This governance is then pushed through the entire organization in terms of setting a consistent security culture, policies, and SETA programs (Alshaikh, 2020; da Veiga et al., 2020; Hu et al., 2012; Sharma & Aparicio, 2022; Van Niekerk & Von Solms, 2010). The apparent absence of such structured governance mechanisms contributed to inconsistent responses and heightened organizational risk.

In cases of major governance failures, leaders may lose credibility and trust. This makes "recovery" more difficult, hampering the ability to correct the root problems (e.g., governance weaknesses and lack of a convenient and secure inter-agency communication tool). Worse, the lessons learned become negative and spread into an organization's culture. These lessons might be more in line with the following:

- Do your immediate job, no matter the security shortcuts or workarounds you need to pull it off.
- Do not speak up, or you will be punished.
- Loyalty to leadership is more important than truth, ethics, following policies and procedures, looking out for stakeholders, doing the right thing, or following relevant laws.
- Expertise and responsibility in your functional area are a liability if they get in the way of a leader's conflicting ambitions.
- Do not admit to mistakes: blame, deflect, point the finger, and if necessary, lie.
- "Security for thee but not for me": Underlings need to follow the rules, but leaders are too important to be constrained by the same.

Organizational environments that internalize the wrong lessons—especially the belief that accountability can be avoided—often suffer morale, particularly among diligent employees who are discouraged from speaking up or punished for following protocol. Declining morale weakens an organization's internal cohesion and increases its cyber risk exposure, as disengaged employees may become more susceptible to human error, internal leaks, or even intentional sabotage. In the wake of the Signalgate incident, reports suggest that internal trust eroded, particularly as leadership figures were perceived as avoiding accountability. According to multiple sources, the administration was resistant to public admissions of fault, contributing to an organizational climate where several top aides resigned or were dismissed, leaving key leaders without a trusted or unified team. Such outcomes can undermine internal and external credibility and reduce an agency's ability to respond effectively to future incidents.

## 4. Discussion

Signalgate is a compelling nation- and organizational-level cybersecurity incident with captivating, often contextual details, which can attract increased interest and attention to the bigger national and organizational cybersecurity issues. Though the actors, stakeholders, organizations, and technologies may change, Signalgate is an archetypal cybersecurity story with patterns and principles repeating throughout virtually every nation and organization. Governments, not-for-profit, and for-profit enterprises spend trillions of dollars annually on security-related initiatives, which have become deeply intertwined with national intelligence, counter-terrorism, military intelligence, cyberwarfare, CTI, and private industry organizational security, where private industry is a primary target of nation-state actors and criminal organizations. Moreover, security researchers have published thousands of peer-reviewed papers, instituted new formal degrees and credential programs, trained thousands of students, presented at hundreds of conferences, and secured billions in grants. Despite this significant global effort and expenditure from the many stakeholders involved, the security landscape is worsening. Globally, we have reached the point where the collective costs of security, including cybercrime, cyberwarfare, and security breaches, were estimated to exceed USD$6 trillion a year in 2021 and are expected to reach as high as USD$10.5 trillion in 2025, which outstrips every country's GDP growth and most countries' inflation rates.[26] Consider that the GDP of the third-largest country in the world, Germany, is approximately USD$4.9 trillion in 2025. Signalgate is thus merely emblematic of the wider problem. To this point, Tony Bradley, senior contributor on cybersecurity at Forbes, opined:

> "The Signalgate scandal, combined with these broader exposures, **reflects a culture problem**.
>
> We often think of cybersecurity as a technical discipline, but most breaches start with human error. Messaging apps like Signal are encrypted and secure — but only if used properly. Platforms like Venmo offer privacy settings — but only if configured correctly. Contact information can be protected — **but only if someone cares enough to do it.**

> **Unfortunately, too many public officials treat digital security as an afterthought — until it becomes a headline.**
>
> What's more frustrating is that these missteps aren't happening in isolation. They're happening among the very people charged with protecting national interests. **If senior government officials are casually sharing classified operations over apps and leaving their digital doors wide open, what hope is there for the rest of us?**" (emphasis added)

Because of similar people and culture issues, what happened in Signalgate can easily occur in any organization. We are on a worldwide cybersecurity trajectory that, if we do not adopt humility, willingness to learn, and willingness to change, adapt, and respond dynamically, we are all "next"—it is only a matter of when, how, who, how bad, and how often. Thus, collective empathy, humility, and a willingness to learn and change are in order because Signalgate is just a symptom of the problem. Despite security debacle after security debacle—public and private—top leadership in global government, not-for-profit organizations, and for-profit firms, large and small, often overlook the crucial nature of national and organizational security, and the role that "tone at the top" plays in the success of a nation or organization in protecting itself against security and privacy breaches. Unless there is a profound cultural shift in how we approach security in research, organizations, practice, and policy, we will likely repeat cycles of breach and remediation without lasting improvement, akin to the Sisyphus metaphor.

Moreover, the Signalgate case serves as a timely reminder that organizational security is only as strong as the weakest link, and invariably, these "weak links" are typically organizational insiders, involving human error (Cram et al., 2019; Crossler et al., 2013; D'Arcy et al., 2009; Galletta et al., 2025; Hu et al., 2012; Johnson & Goetz, 2007; Johnston & Hale, 2009; Johnston et al., 2015; Lowry et al., 2023; Lowry & Moody, 2015; Lowry et al., 2017; Posey et al., 2013). These human issues run a spectrum and have different predictors, explanations, and solutions, and thus vex organizations that sometimes prioritize technical solutions without commensurate attention to organizational and behavioral factors, which can be undermined by human error. These often overlooked organizational and behavioral cybersecurity issues are the Achilles heel of all organizations—even those sometimes viewed as "too big to fail," such as the US Government's trillions of dollars invested in cybersecurity. We argue that a nation or firm can waste significant resources by investing in robust cybersecurity technologies and protocols, without equally investing in and taking seriously the underlying organizational issues of governance, leadership, organizational structures, and the human elements of cybersecurity.

These vexing issues range from serious personal problems outside of work (bankruptcy, divorce, death of a family member, lawsuits, personal health issues, financial distress, severe mental health crisis), fatigue, and emotional stress that distract employees from engaging in cybersecurity threat awareness and prevention, and open them up to mistakes and external exploits (Burns et al., 2019; Cram et al., 2021; D'Arcy & Lowry, 2019; D'Arcy et al., 2014; Mehrizi et al., 2022); poor leadership and governance that undermine cybersecurity and organizational morale (Aksoy, 2025; Armour, 2017; Galletta et al., 2025; Gillespie et al., 2014; Oluka, 2023); lack of measurement and incentives that reward employees for engaging in protective measures, or conflicts and contradictions among these, undermining organizational security (Bada et al., 2019; Chaudhary, 2024; Jacobs et al., 2023; Li et al., 2022); lack of a sense of accountability and organizational commitment in organizations (Posey et al., 2015; Posey et al., 2013; Vance et al., 2015); low employee motivation to protect organizational assets (Lowry & Moody, 2015; Posey et al., 2015; Posey et al., 2013; Willison & Lowry, 2018; Xu et al., 2024); poorly designed and delivered security policies that often backfire (Lowry et al., 2023; Lowry & Moody, 2015; Lowry et al., 2015); heavy-handed sanctions or threats from management that undermine morale and motivation to protect the organization (Burns et al., 2023; D'Arcy et al., 2009; Herath & Rao, 2009; Willison et al., 2018); work role conflicts or poor work design that encourages the use of Shadow IT and workarounds (Alter, 2014; Haag & Eckhardt, 2017; Silic & Back, 2014); poor training and lack of understanding/awareness among employees regarding security hygiene and relevant threats (Ahmad et al., 2021; D'Arcy et al., 2009; Hull et al., 2023; Khando et al., 2021; Silic & Lowry, 2020; Tsohou et al., 2015); and the fundamental lack of psychological capital, self-efficacy, and response-efficacy for employees to effectively recognize and address organizational security threats (Boss et al., 2015; Burns et al., 2017; Gwebu et al., 2018; Herath & Rao, 2009; Hull et al., 2023; Johnston et al., 2015; Knight & Nurse, 2020; Lowry et al., 2023; Sarkar & Shukla, 2023; Schuetz et al., 2020).

In addition, in virtually every organization, a non-trivial subset of employees may engage in policy-violating or unlawful behavior who slip through HR screenings, come under the influence of criminal organizations or nation-state actors, or may become involved in misconduct during the course of their employment, often spurred by major personal issues and life changes. Cyber criminals, international cybercrime gangs and syndicates, and sophisticated nation-state actors frequently use techniques to turn otherwise law-abiding employees into criminals who can wreak havoc on their organizations: digital grooming and rapport building (Lorenzo-Dus, 2022; Sarkar & Shukla, 2023); bribery and side payments (Bada & Nurse, 2023); gig-economy recruiting turned into economic coercion;[27] human trafficking and threats to legal residency or legal work status (Lazarus et al., 2025); blackmail and extortion (Bada & Nurse, 2023); sexual honey pots (real world or virtual) and sextortion (Ray & Henry, 2025); disinformation and misinformation campaigns (Bada & Nurse, 2023; Guo et al., 2021); diaspora and identity exploitation (typically aiming at ethnicity, national identity, sexual identity, religion, or movements that evoke passionate responses) (Homoliak et al., 2019), and more. These insidious activities open up the most dangerous possibilities of insiders using their privileged inside access to engage in harmful acts for revenge or retaliation, curiosity or thrill-seeking, financial gain, fraud, extortion, political agenda, and religious causes (Bossler & Berenblum, 2019; Dupont et al., 2024; Willison & Lowry, 2018; Willison et al., 2018; Wu et al., 2023). There have already been bigger, more dangerous cybersecurity cases, like those between Edward Snowden and the NSA, that dwarf Signalgate in magnitude and damage (Edgar, 2017; Lucas, 2014).

Despite what we know in research and practice, organizations have yet to receive the "signal" and instead have often heavily focused on investing in technical solutions while overlooking the more critical organizational, governance, and human factors at the root cause of their organizational security issues. Until this significant organizational imbalance in security root causes and solutions substantially changes, organizational security breaches and exploits will only worsen, and trillions of dollars will be wasted that could have been used to provide greater value for stakeholders. This is no minor consideration: with trillions of dollars already being spent and growth in costs outstripping inflation, organizational and national budgets are at stake. Recently, NATO nations have agreed to expend 5% of their GDP on national security (nearly doubling existing expenditures, despite stagnating growth and revenues), of which approximately 1.5% of GDP of this is directly related to cybersecurity; for comparison, 6.7% of Russia GDP is spent on defense, and 7.2% of China's GDP is spent on defense, with estimated larger portions going to cybersecurity.[2] We are at a global tipping point where the current technical-first security approach risks becoming economically unsustainable and creating broader organizational vulnerabilities.

What we have globally chosen, predominantly a technology-first approach to national security and cybersecurity, is akin to the curse of Sisyphus, who was eternally condemned to push a boulder up a hill, only to have it roll back down in an endless cycle. Each cycle becomes more challenging, and relying solely on resources and effort has not prevented repeated setbacks. Unlike the myth, this challenge directly affects billions of people. Our cybersecurity crisis has consequences for billions of people, and is chosen by human national and organizational leadership. Only these human leaders, their constituents, and stakeholders can disrupt the doom loop through our collective choices to break the chain of cybersecurity organizational failure.

However, there is hope, and there are meaningful actions that nations and organizations can choose to opt out of this cycle. We used the NIST CSF framework to provide lessons that can be applied to virtually any organization, and we see the major takeaways as follows, which we also believe are the most pressing opportunities for researchers to focus on, as well:

- Strong security leadership through "tone at the top" is fundamental. Organizational security efforts must be understood and championed non-hypocritically from the Board, CEO, CISO, C-level executives, and management, and embedded in an organization's ethos, culture, and all employees.
- Zero-trust architecture and principles of least privilege are persuasive governance concepts that must be engineered into every level and aspect of an organization, with no exceptions.
- Confusing signals about chain-of-command, authority, and accountability undermine organizational governance and security. Roles, authority, and responsibilities must be clearly articulated. Leaders must be held accountable for all security aspects, or the wrong signal is sent to the organization.
- Organizations "get" what they incentivize and what they measure. All aspects of key security practices to protect an organization's assets must be tied to clear measures and incentives for all levels of employees in the organization.
- The concepts of independence, oversight, transparency, auditing, and supporting whistle-blowing must be fiercely defended for any organization to achieve effective security governance. The independence and expertise of the Board of Directors, CISO, CFO, external auditors, and external regulators must be fully supported.
- Mergers and acquisitions, and other times involving major organizational transition or shock (e.g., massive layoffs), should be considered especially high-risk events that exponentially increase the attack surface and vulnerabilities of an organization, and thus require additional planning, governance, analysis, and effective communication and measures to avoid the increased vulnerabilities.

---

[2] John R. Deni and Ryan Arick (2025). "What counts as 'defense' in NATO's potential 5 percent spending goal?," New Atlanticist (published June 20, 2025; last accessed June 24, 2025).

- Hiring and vetting employees, contractors, and service providers are core security vulnerability points requiring consistent and careful screening. Consistent standards, assessments, audits, and monitoring should be applied, depending on the level of responsibility and organizational data access.
- Organizations need to improve their handling of their meeting and communication cultures and practices if they are going to improve their security postures. Aside from the inefficiencies that occur from "all-hands-on-deck" meetings and meetings that are primarily informational and involve people "just in case," and the onslaught of emails with too much information to too many people, organizations need to tie meetings and communication to security principles better, meaning: have a focused agenda for all meetings, only invite appropriate and required roles, understand how, when, and with what technologies should be used for communication; rethink and better govern how employees communicate about what, and challenge prevailing toxic meeting, email, chat, and Zoom practices.
- Organizational insiders can either be allies in security or the weakest links that lead to the security problem. The only way to foster allies is through the above points, providing training, culture, support, and a learning culture that allows people to make and address mistakes.

#### 4.1.1. Conclusion

When it comes to organizational security, we do not have the luxury of picking and choosing the known and unknown threats that will apply to us or the consequences, even and especially when they could pose extreme inconvenience to one's leadership agenda, organizational strategy, product launch, mergers and acquisitions, quarterly earnings, public relations, and so on. Leaders cannot explain organizational security issues as "IT's problem, not our problem." We only have the choice of how we will mitigate and address the known and unknown risks, and the potential consequences we are willing to bear from our choices. Leaders too often forget that they will bear a personal cost of these breaches and so will members of their organization, if not the entire organization. Firing a scapegoat or two will not absolve them of the consequences. Once these breaches occur, the consequences directly affect both leaders and their organizations, with profound implications for reputation and stability. If not immediately addressed, problems can escalate, leading to greater disruption and unintended negative consequences, often worse than the initial impact. Attempts to minimize, deny, or reframe the incident afterward do not resolve the underlying issues and can compound the harm. It takes courage and leadership to confront the problem directly, admit the failing, right the wrongs, and repair trust. Otherwise, leaders risk losing credibility, authority, and stakeholder trust, leaving the organization without clear direction or effective governance.

## Acknowledgements

We first express deep appreciation to Nanyang Technological University's (Singapore) Nanyang School of Business for the opportunity for the first author to develop the initial Signalgate case and to create and teach a new organizational security course for managers and executives during Spring 2025 for its esteemed full-time MBA program, in which this case was first delivered and developed. We appreciate the funding, housing, infrastructure, and support for a one-year fully funded sabbatical with full salary, research funds, housing, office, software, and computing infrastructure support for the first author that made this research possible at Nanyang Technological University. Accordingly, we especially recognize from NTU: (1) Prof. Waifong Boh (President's Chair in Information Systems; Associate Provost; Co-Director for NTU Centre in Computational Technologies for Finance (CCTF) with the College of Engineering; Director of the Information Management Research Centre (IMARC); and Vice President Lifelong Learning & Alumni Engagement); (2) Prof. Jun Yang (Dean of the Nanyang School of Business; President's Chair in Finance); and (3) Prof. Anandasivam "Anand" Gopal (President's Chair in Information Systems and Innovation and Division Chair of the Division of Information Technology and Operations Management [ITOM]). We are also grateful to the similar financial and infrastructure support from Virginia Tech and the Pamplin College of Business for this one-year sabbatical at NTU, as well as the support and contributions of VT's Center for Security, Privacy, and Trust and VT's Integrated Security Education and Research Center (ISERC). From VT's leadership, we especially want to acknowledge the outstanding support of: (1) Prof. Cyril Clarke (VT Executive Vice President and Provost); (2) Prof. Saonee Sarker (Richard E. Sorensen Chair and Dean of the Pamplin College of Business); and (3) Dr. Quinton J. Nottingham (Department head for the Business Information Technology Department within the Pamplin College of Business).

We also express gratitude for constructive feedback from Anthony Vance, [**TO BE COMPLETED**].

We finally express gratitude for the many anonymous constructive comments we received from Nanyang School of Business MBA students, global industry leaders, and global academics.

---

[1] Throughout our case analysis, we refer to relevant NIST CSF controls that were violated in bolded superscript, for example: **[ID.GV-2]**. These NIST CSF controls relate to how organizations are supposed to address specific NIST CSF functions (e.g., ID categorizes the family of controls for the Identify function.

[2] The Wall Street Journal reported that Waltz had hosted Signal group chats with Cabinet members on a number of topics: McGraw, Meridith; Ward, Alexander; Dawsey, Josh (2025), "Mike Waltz Is Losing Support Inside the White House," The Wall Street Journal (published March 30, 2025; last accessed September 8, 2025).

Politico followed up with a report from anonymous insiders that Waltz had set up at least 20 such groups: Dasha Burns (2025), "Waltz's team set up at least 20 Signal group chats for crises across the world," Politico (published April 02, 2025; last accessed September 08, 2025).

Finally, the New York Times reported that Hegseth had set up a Signal group chat titled "Defense | Team Huddle" that included his wife, brother, and a dozen other people, where he allegedly shared attack details with this group: Greg Jaffe, Eric Schmitt, and Maggie Haberman (2025). "Hegseth Said to Have Shared Attack Details in Second Signal Chat," The New York Times (published April 20, 2025; last accessed September 08, 2025).

[3] BTAC (2022). "Leadership transitions: Impacts on organizational change," The Behavioral Threat Analysis Center (BTAC) (published 2022; last accessed May 2, 2025). The BTAC is a unit of the Department of Defense's Insider Threat Management and Analysis Center (DITMAC), operating under the Defense Counterintelligence and Security Agency (DCSA).

[4] For example, in a survey of 720 executive, the IBM found mergers and acquisitions (M&As) to be high-value targets with one and three firms undergoing M&As suffering breaches: IBM (2020). "Assessing cyber risk in M&A: Unearth hidden costs before you pay them," IBM Institute for Business Value (last accessed May 2, 2025).

[5] David Smith (2024). "Warren says Trump's 'unprecedented' actions during transition risk security," The Guardian (published November 22, 2024; last accessed September 08, 2025).

[6] Alice Miranda Ollstein (2024). "Trump's transition is happening over private emails. Federal officials are nervous," Politico (published December 18, 2024); last accessed September 08, 2025).

[7] Wikipedia (2025), "2025 dismissals of U.S. inspectors general," Wikipedia (last accessed September 08, 2025).

[8] Meryl Kornfield and Lisa Rein (2025). "Trump oversight picks include scandal-hit ex-lawmaker, antiabortion lawyer," Washington Post (published May 29, 2025; last accessed September 08, 2025).

[9] KPMG (2014), "Fraud risk management: Developing a strategy for prevention, detection, and response," (published May 2014; last accessed June 23, 2025).

[10] Emma Colton (2025). "Biden-era guidance encouraged use of Signal app by highly-targeted govt officials: 'Best practice,'" Fox News (published March 25, 2025; last accessed September 08, 2025).

[11] Anna Mulrine Grobe (2025). "Why security officials keep using the Signal app despite risks," The Christian Science Monitor (published May 08, 2025; last accessed September 08, 2025).

[12] Margaret Hartmann (2025). "Mike Waltz: 'We're Trying to Figure Out' How I Did This," New Yorker (published March 26, 2025; last accessed June 23, 2025.

[13] R. Barnes, B. Beurdouche, R. Robert, J. Millican, E. Omara, and K. Cohn-Gordon (2024). "The Messaging Layer Security (MLS) Protocol," Request for Comments (RFC) 9420, Internet Engineering Task Force (IETF), Standards Track, ISSN: 2070-1721 (last updated July 07, 2024; last accessed May 2, 2025).

[14] Alexandra Hutzler (2025). "What to know about classified information as Signal chat fallout continues," ABC News (published March 28, 2025; last accessed June 23, 2025).

[15] Dell Cameron (2025). "Here's What Happened to Those SignalGate Messages," Wired (published April 15, 2025; last accessed September 08, 2025).

[16] David DiMolfetta (2025). "Government interest in chat archiving service skyrockets following Signalgate," NEXTGOV/FCW (published March 31, 2025; last accessed September 8, 2025).

[17] Ronny Reyes (2025). "Trump envoy Steve Witkoff did not use Signal while in Russia, White House says — as fears grow over Kremlin hacking of app," New York Post (published March 26, 2025; last accessed June 23, 2025).

[18] Sean Michael Newhouse (2025). "Senior Democrat wants to know who in the Trump White House got Top Secret clearance without a complete background check," Government Executive (published January 31, 2025; last accessed September 08, 2025).

[19] Lara Seligman, Vera Bergengruen, and Nancy A. Youssef (2025). "Trump Team Blindsided by Details of Sexual-Assault Allegation Against Hegseth," The Wall Street Journal (published November 21, 2024; last accessed September 08, 2025).

Ariana Baio (2025). "Trump's transition team was 'blindsided' by new details of Pete Hegseth's sexual-assault allegation," The Independent (published November 22, 2025; last accessed September 08, 2025).

[20] Chad de Guzman (2025). "Hegseth Comes Under Fire Amid Reports of Sharing Sensitive Info in Another Signal Group Chat," Time (published April 21, 2025; last accessed June 23, 2025).

[21] Ray Fernandez (2025). "Secret US war plans leaked in a Signal group chat by accident," Moonlock (published March 27, 2025; last accessed September 08, 2025).

Tanner Stening (2025). "White House Signal mishap 'would never have happened' if leaders used official channels, cybersecurity expert says," Northeastern Global Times (published March 25, 2025; last accessed September 08, 2025).

[22] James LaPorta (2025). "NSA warned of vulnerabilities in Signal app a month before Houthi strike chat," CBS News (last updated March 25, 2025; last accessed April 04, 2025).

[23] Ronald J. Deibert (2025). "The Real Lesson of Signalgate: A Surveillance Arms Race Has Poked a Gaping Hole in National Security," Foreign Affairs (published April 24, 2025; last accessed June 23, 2025).

[24] Security (2025). "Lack of visibility is the biggest challenge for security leaders when safeguarding digital communications," Security (published July 13, 2021; last accessed September 08, 2025).

[25] Associated Press (2025). "Hegseth had an unsecured internet line set up in his office to connect to Signal, AP sources say," The Economic Times (published April 25, 2025; last accessed June 23, 2025).

[26] Nivedita James Palatty (2025). "90+ Cyber Crime Statistics 2025: Cost, Industries & Trends," Astra (last updated February 6, 2025; last accessed April 5, 2025)

[27] UNODC (2024). "Transnational Organized Crime and the Convergence of Cyber-Enabled Fraud, Underground Banking and Technological Innovation in Southeast Asia: A Shifting Threat Landscape," United Nations Office on Drugs and Crime (UNODC) (published October 2024; last accessed June 25, 2025).